# Kelvin-Helmholtz instability in an ultra-thin air film causes drop splashing on smooth surfaces

Yuan Liu *, Peng Tan * †1 and Lei Xu * 1

*Department of Physics, The Chinese University of Hong Kong, Hong Kong, China, and †State Key Laboratory of Surface Physics, Department of Physics, Fudan University, Shanghai 200433, China

When a fast-moving drop impacts onto a smooth substrate, splashing will be produced at the edge of the expanding liquid sheet. This ubiquitous phenomenon lacks a fundamental understanding. Combining experiment with model, we illustrate that the ultra-thin air film trapped under the expanding liquid front triggers splashing. Because this film is thinner than the mean free path of air molecules, the interior air flow transfers momentum with an unusually high velocity comparable to the speed of sound, and generates a stress ten times stronger than the common air flow. Such a large stress initiates Kelvin-Helmholtz instabilities at small length scales and effectively produces splashing. Our model agrees quantitatively with experimental verifications, and brings a fundamental understanding to the ubiquitous phenomenon of drop splashing on smooth surfaces.

drop impact | splash | thin air film | Kelvin-Helmholtz instability

The common phenomenon of drop splashing upon impacting on smooth surfaces may seem simple and natural to most people, however its understanding is surprisingly lacking. Splashing is crucial in many important fields, such as the sprinkler irrigation and pesticide application in agriculture, ink-jet printing and plasma spraying in printing and coating industries, and spray cooling in various cooling systems; therefore its better understanding and effective control may make a far-reaching impact on our daily life. Starting from nineteenth century, extensive studies on drop impact and splash have covered a wide range of control parameters, including the impact velocity, drop size, surface tension, viscosity, drop composition and substrate properties [1–12], and various splashing criteria have been proposed and debated [13–18]. Nevertheless, at the most fundamental level the generation mechanism of splashing remains a big mystery.

Recently a breakthrough has surprisingly revealed the importance of surrounding air, and suggested the interaction between air and liquid as the origin of splashing [15, 19, 20]. However, this interaction is highly complex: below the drop air is trapped at both the impact center and the expanding front [21–34], and above it the atmosphere constantly interacts with the top interface. As a result, even the very basic question of which part of air plays the essential role is completely unknown. Moreover, the analysis from classical aerodynamics [18] indicates that the viscous effect from air totally dominates any pressure influence, while the experiment contradictorily revealed a strong pressure dependence [15]. Even more puzzling, it was revealed that the speed of sound in air plays an important role in splashing generation [15], although the impact speed is typically 10 to 100 times slower! Therefore, an entirely new and non-classical interaction, which can directly connect these two distinct time scales, is required to solve this puzzle. Due to the poor understanding of underlying interaction, the fundamental instability that produces splashing is unclear: the prevailing model of Rayleigh-Taylor (RT) instability [35] contradicts with the pressure-dependent observation [15, 19, 20]; while the recent proposition of Kelvin-Helmholtz (KH) instability lacks direct verification [19, 36]. Therefore, clarifying the underlying air-liquid interaction and further illustrating the splash-generating instability mechanism are currently the most critical issue in the field.

To tackle this issue, we fabricate special porous substrates that enable effective air drainage at carefully-designed locations, and systematically probe the air-liquid interaction and the splash-generating instability. By making pores at either the impact center or the expanding edge, we reveal that the air trapped under the expanding edge triggers splashing. Because the trapped air is thinner than the mean free path of air molecules, the interior air flow transfers momentum with an unusually high velocity comparable to the speed of sound, and generates a stress ten times stronger than the common air flow. Such a large stress initiates Kelvin-Helmholtz instabilities at very small length scales and effectively produces splash. Our model agrees quantitatively with experimental verifications, and brings a fundamental understanding to the ubiquitous phenomenon of drop splashing on smooth surfaces.

## Results

We release millimeter-sized liquid drops from various heights and impact them onto different substrates. To guarantee reproducible and pronounced splash, we choose liquids with low surface tensions. The liquids are also in the low-viscosity regime where surface tension dominates the viscous effect [19,20]. Three types of substrates are used: smooth substrates, patterned leaking substrates, and patterned non-leaking substrates, as shown in Fig.1a. For patterned substrates made by

---

**Significance**

Liquid drops always splash when they impact smooth surfaces with large enough speeds. This common phenomenon is crucial in many important fields such as agriculture, printing, surface coating, and spray cooling. However, despite extensive studies over one century, the origin of splashing remains a big mystery. Combining experiment with model, we show that the air trapped under the liquid drop forms a special flow within a nanoscale gap. This air flow produces a stress ten times stronger than the common air flow, and generates small Kelvin-Helmholtz instabilities that trigger splash. Our model agrees quantitatively with the experimental verifications and brings a fundamental understanding to the general phenomenon of drop splashing on smooth surfaces.

Reserved for Publication Footnotes

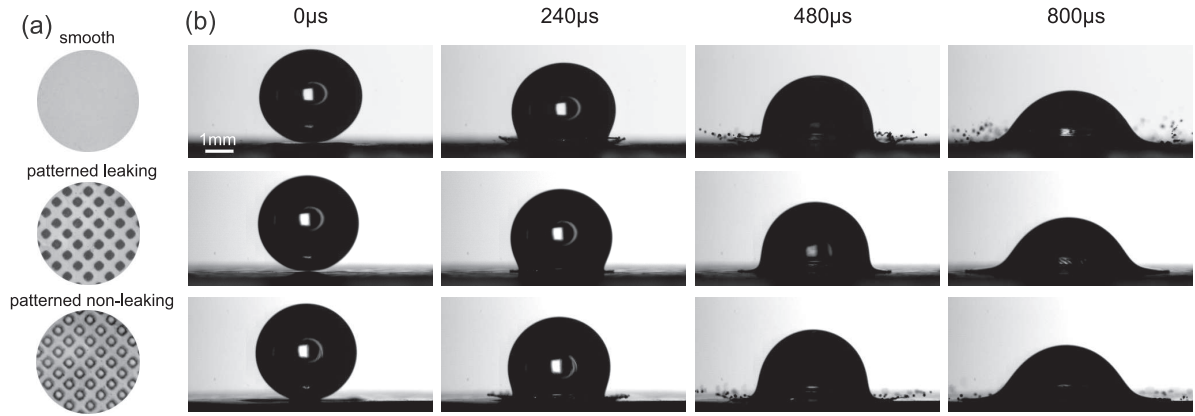

**Fig. 1.** Distinct splashing outcomes for liquid drops impacting on different substrates. (a) Images of the three different substrates: smooth, patterned leaking and patterned non-leaking substrates. The pore diameter and the spacing between pores are both $75 \pm 5\,\mu m$. The leaking and non-leaking substrates have identical patterns of pores, except that for the former substrate pores are all the way through while for the latter one they are only half-way through. (b) Corresponding splash outcomes for the three types of substrates, for an ethanol drop with diameter $3.5 \pm 0.1$mm and impact velocity $V_0 = 1.92 \pm 0.01$m/s. Splashing occurs significantly on the smooth substrate, disappears completely on the patterned leaking substrate, and reappears on the patterned non-leaking substrate. It clearly demonstrates that the air entrainment *under* the drop causes splashing.

optical lithography, the diameter of pores is $75 \pm 5\,\mu m$, much smaller than the millimeter-sized liquid drops. The leaking and non-leaking substrates have identical patterns of pores, except that for the former case pores are all the way through while for the latter case pores are only half-way through. Thus the leaking substrate reveals the outcome for impacts with effective air drainage; while the non-leaking substrate provides the zero-leakage comparison. All experiments are performed under the atmospheric pressure, $P_0 = 101$kPa, and recorded by high-speed photography. To make sure that our results are generally valid, we perform experiments with six different liquids, two different substrate materials and various impact velocities (see Supplementary Information). All experiments exhibit consistent behaviors which demonstrate the robustness of our finding.

In Fig.1b we show the corresponding impact outcomes for the three substrates shown in Fig.1a, at the same impact velocity $V_0 = 1.92 \pm 0.01$m/s (also see the movie S1). Apparently, splashing occurs significantly on the smooth substrate, disappears on the patterned leaking substrate, and reappears on the patterned non-leaking substrate (top, middle and bottom rows respectively). The complete disappearance of splashing on the leaking substrate unambiguously proves that the air trapped *under* the liquid causes splashing, and when it drains away splashing vanishes. Furthermore, the reappearance of splashing in the third row of patterned but non-leaking substrate confirms once again that it is the air drainage instead of the surface pattern that eliminates the splashing.

More specifically, air is trapped under the liquid at two distinct locations: the impact center [21–32] and the expanding edge [33], which are separated by a large wetted region in between (see Fig.2c). Which entrapment is essential for splashing? We tackle this question with impact experiments on two corresponding substrates as shown in Fig.2a: the top substrate enables a complete drainage of air entrapment at the impact center, while the bottom substrate only eliminates air entrapment at the edge. Great care is taken to make sure that the initial contact always occurs at the substrate center. The impact results are demonstrated in Fig.2b (also see the movie S2): apparently draining air at the center does not eliminate splashing (the top row); while removing air entrainment at the edge eliminates splashing completely (the bottom row). This finding clearly indicates that it is the air trapped under the expanding edge [33] that plays the essential role.

By making pores at different regions, we clarify that splashing is created by the air entrainment under the expanding front. Next we illustrate the detailed air-liquid interaction within this air entrainment. According to the previous experiment [33], this entrainment is an ultra-thin air film with a typical thickness of $10 \sim 100$nm, less than or comparable to the mean free path of air molecules (about 70nm at $P_0 = 101$kPa). As a result, the continuous aerodynamics breaks down and the microscopic picture in the Knudsen regime must be considered. Inside this film, the air molecules right below the liquid surface naturally obtain an average velocity identical to the expanding liquid front, $V_e$, and then transfer this momentum to the nearby solid surface $10 \sim 100$nm away (see the relevant geometry and quantities drawn in Fig.2c, d). Because the travel distance is smaller or comparable to the mean free path, the air molecules essentially reach the solid surface with *ballistic* motions, which have the velocity comparable to the speed of sound. Such a fast motion enables a surprisingly high efficiency in momentum transfer, and produces a large stress involving the speed of sound. A detailed calculation by P. G. de Gennes gives the exact expression of the stress [37]: $\Sigma_G = \rho_a \cdot c \cdot V_e / \sqrt{2\pi\gamma}$, with $\rho_a$ the density of air (the value at $P_0$ is used because this air entrainment is directly open to the outside atmosphere as shown in Fig.2d), $c$ the speed of sound in air, $V_e$ the expanding velocity of the liquid front, and $\gamma = 1.4$ the adiabatic gas constant. Therefore, the expression of $\Sigma_G$ based on the ballistic motion of air molecules in the ultra-thin air film naturally connects the two distinct velocities, the speed of sound and the expanding velocity, and explains the outstanding puzzle previously observed [15].

We further illustrate $\Sigma_G$ by comparing it with the Bernoulli stress from a wind blowing across a liquid surface under common circumstances. For the common air flow, the stress takes the Bernoulli expression of $\rho_a \cdot V_e^2$, which differs from $\Sigma_G \sim \rho_a \cdot c \cdot V_e$ by a typical factor of $c/V_e$. Plugging in the characteristic values of $c \sim 100$m/s and $V_e \sim 10$m/s, clearly $\Sigma_G$ is larger than the common situation by one order of magnitude, and thus behaves as a special air flow ten times stronger. We propose that this special air flow can initiate Kelvin-Helmholtz (KH) instabilities around the liquid front and produce splash-ing, as schematized in Fig.2d (the dashed curve indicates the instability, drawn not on scale).

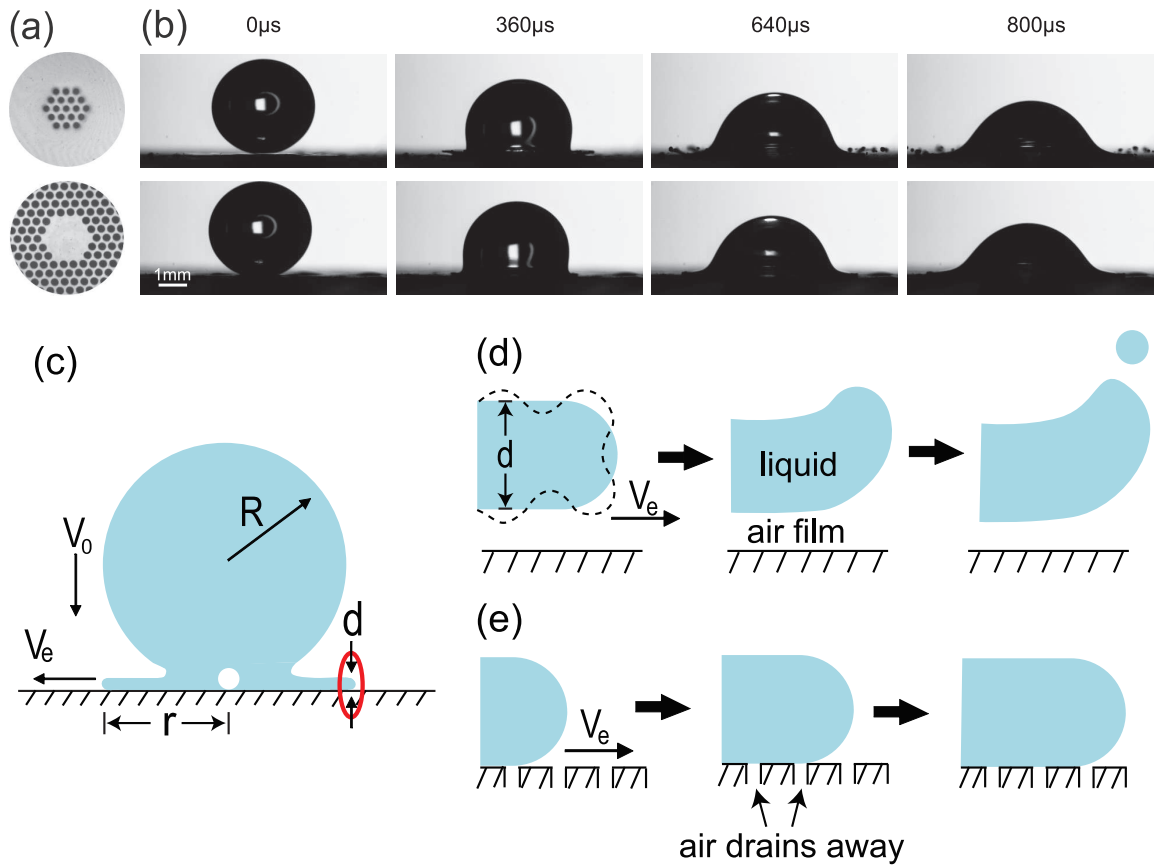

**Fig. 2.** The air entrapment at the expanding edge causes splashing. (a) Images of the two substrates with leaking areas at either the center or the edge. The pores have the diameter $75 \pm 5\,\mu m$. The top substrate enables a complete drainage of air entrapment at the impact center, while the bottom substrate eliminates air entrapment at the edge. (b) Corresponding splashing results on these two substrates, for an ethanol drop with impact velocity $V_0 = 1.92 \pm 0.01 m/s$ and diameter $3.5 \pm 0.1 mm$. Splashing occurs in the first row but disappears in the second, revealing that the air trapped at the expanding edge causes splashing [33]. (c) A schematics showing the impact geometry and relevant quantities: the impact velocity is $V_0$, the front expanding velocity is $V_e$, the expanding radius is $r$, and the liquid sheet thickness is $d$. Air is trapped under both the center and the expanding front, but the analysis is focused only at the expanding front (within the red ellipse). (d) Cartoon pictures (drawn not on scale) demonstrating the detailed splashing process on a smooth substrate. The ultra-thin air film trapped under the liquid initiates the KH-instability, as indicated by the dashed curve, which subsequently develops into splashing. (e) The corresponding situation on a leaking substrate. No air film exists because air drains away, which consequently eliminates the instability and splashing.

To obtain a quantitative understanding, we construct a KH-instability model inside the ultra-thin air film. Following the classical work by John W. Miles [38], we write the differential equation for the interface according to stress balance:

$$L\eta + m\frac{d^2\eta}{dt^2} = -p_a \qquad [1]$$

Here $\eta = ae^{i(kx-\omega t)}$ is the small-amplitude disturbance at the interface, $L$ is a linear operator such that $L\eta$ gives the stress resisting a deformation $\eta$ of the surface, $m = \rho_l/k$ is the effective liquid mass per unit area with $\rho_l$ the liquid density, and $p_a$ is the aerodynamic stress acting on the interface. Apparently this equation is equivalent to the Newton's 2nd law $F = ma$ and thus should be generally valid. For a length scale much smaller than the capillary length ($\sim 1mm$), the gravity can be completely neglected and only surface tension matters, which leads to $L\eta = \sigma k^2 \eta$, with $\sigma$ being the surface tension coefficient of the liquid [38, 39]. In particular, the aerodynamic stress can be expressed as $p_a = -\Sigma_G k\eta$, where we have replaced the Bernoulli stress in the original literature with $\Sigma_G$ and the minus sign corresponds to the KH-instability [38].

Plugging in all these terms to Eq.[1] leads to the dispersion relation:

$$\omega^2 = (\sigma k^3 - \Sigma_G k^2)/\rho_l \qquad [2]$$

Note that the dispersion relation is time dependent because $\Sigma_G \propto V_e$ varies with time. The system will go unstable once the right hand side becomes negative, which happens when the destabilizing stress $\Sigma_G$ overcomes the stabilizing effect from $\sigma$. By taking $d\omega/dk = 0$, we obtain the wavenumber of the most dangerous mode that grows the fastest, $k_m = 2\Sigma_G/3\sigma$. Plugging $k_m$ back into Eq.[2] we get $\omega_m^2 = -\frac{4\Sigma_G^3}{27\sigma^2 \rho_l}$ and thus the growth rate of the most dangerous mode is $|\omega_m| = \sqrt{\frac{4\Sigma_G^3}{27\sigma^2 \rho_l}}$. With a typical measurement from experiment, $V_e = 6m/s$, we can estimate the numerical values of the length and time scales for the most dangerous mode: $k_m^{-1} \sim 40\,\mu m$ and $|\omega_m|^{-1} \sim 60\,\mu s$. Both values are much smaller than the conventional KH-instability situations. $k_m^{-1} \sim 40\,\mu m$ also agrees well with the size of secondary splashing droplets.

Due to the unusually large stress of $\Sigma_G$, which is ten times stronger than the conventional Bernoulli expression, the air flow in an ultra-thin air film can generate KH-instabilities with an unusually small length scale, $k_m^{-1} = 3\sigma/2\Sigma_G$. The underlying physics is rather straightforward: a gentle breeze can generate slowly-varying long-wavelength disturbances, while a strong wind may produce much smaller agitations; the exact size depends on the stress balance between the aerodynamic stress, $\Sigma_G$, and the restoring effect from the surface tension, $\sigma$. Because the large $\Sigma_G$ causes a small $k_m^{-1}$, tiny undulations that can rarely be generated by a regular air flow may now show up.

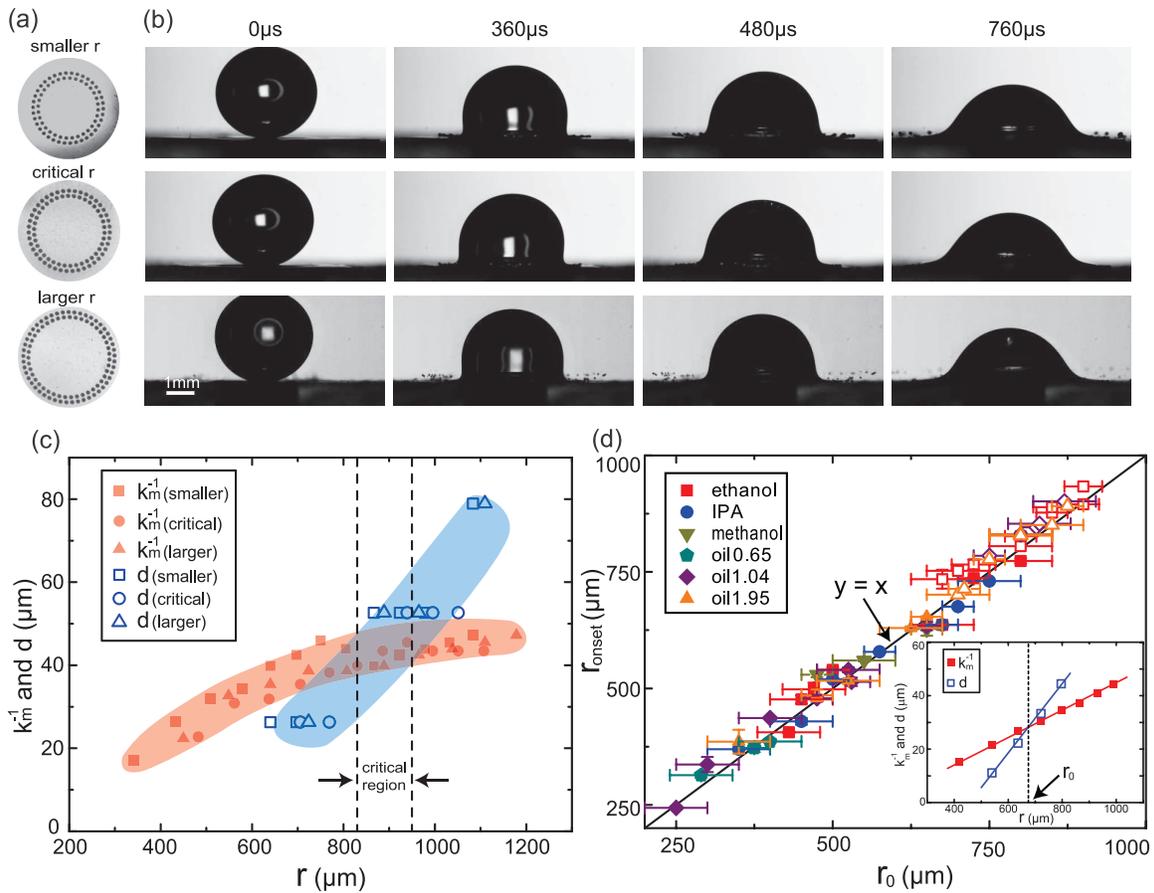

**Fig. 3.** The match between the instability size and the liquid sheet thickness initiates the splash. (a) Three substrates with leaking areas around three different radii: smaller, critical and larger r's. (b) Corresponding splashing outcomes at $V_0 = 1.92$ m/s on these three substrates. Significant splashing appears in the top and bottom rows, but no splashing appears in the middle row. Apparently there exists a critical radius around which the air disturbance is most crucial, and the air drainage there completely eliminates splashing. (c) The critical region overlaps with the resonance location $r_0$, where $k_m^{-1} = d$. We determine $r_0$ by intersecting $k_m^{-1}$ and $d$ curves from independent measurements (note that $d$ saturates around $50\,\mu\text{m}$ without instability but thickens to $80\,\mu\text{m}$ with instability). The critical region determined by the leaking substrate locates between the two dashed lines. The nice overlap between the curve intersection and the critical region agrees well with our model. (d) The splashing onset locations measured from experiments versus the values predicted by our model on smooth substrates. Inset: finding the onset location, $r_0$, with the model. Main panel: for various liquids, substrates and impact velocities, the experimental measurement, $r_{onset}$, agrees excellently with the model prediction, $r_0$. The close and open symbols distinguish the two substrates, glass and optical adhesive NOA81; and different shapes indicate different liquids.

More interestingly, there is another intrinsic length scale in this problem: the thickness of the liquid sheet, $d$. Thus at a specific moment, the instability size $k_m^{-1}$ may match the thickness $d$ which generates a spatial 'resonance' in length scales. We propose that it is this spatial resonance that significantly boosts the growth of the instability and causes splashing, as illustrated in Fig.2d. By contrast, on a leaking substrate the ultra-thin air film does not exist due to the drainage of air, which consequently eliminates the small KH-instability and the splashing, as demonstrated in Fig.2e.

Our model essentially describes a resonance in length scales, $k_m^{-1} = d$, which strongly enhances the KH-instability and produces splashing. Since both $k_m^{-1}$ and $d$ can be measured experimentally, we can quantitatively test this model. In particular, because both $k_m^{-1}$ and $d$ vary with time as the liquid front advances, their match must occur at a specific location, $r = r_0$, where the disturbance from air should be most critical. To explore this critical location, we fabricate various substrates with leaking regions around specific radii, as shown in Fig.3a. The corresponding impact outcomes are demonstrated in Fig.3b: splashing disappears completely when pores are made around a critical radius, as illustrated by the middle row; while significant splashing occurs when pores are made at slightly smaller or larger radii, as shown by the top and bottom rows (also see movie S3). Apparently, there does exist a critical radius around which the disturbance from air is most critical, and the effective air drainage at this location can completely eliminate or significantly reduce splashing. The existence of a critical location agrees well with our resonance picture.

We further verify that exactly at this critical location the resonance condition, $k_m^{-1} = d$, is satisfied. Without any fitting parameter, we directly obtain $k_m^{-1}$ and $d$ from independent measurements, and plot them as close and open symbols in Fig.3c respectively. By definition, these two sets of data intersect at $r_0$ where $k_m^{-1} = d$; in addition, the critical region identi-fied by the leaking substrate in Fig.3a is indicated between the two dashed lines. Clearly the data intersection and the critical region between dashed lines overlap quite nicely, verifying that the critical region for splashing is indeed the resonance location, $r_0$, predicted by our model. More experiments with different liquids and velocities confirm that the splashing criterion, $k_m^{-1} = d$, is robust and universal (see the Supplementary Information, Fig. SI-1a, b). This splash criterion also satisfactorily explains the empirical relation observed in the previous experiment [15]. We further clarify that although the surface tension dominates the viscosity in stabilizing the system, the liquid viscosity does plays a role in the model, through its strong influence on $V_e$ and $d$ [15, 19].

More generally, the capability of identifying $r_0$ with our model enables the quantitative prediction of the precise location, where splashing should first appear, on any smooth substrate. From the experiments on leaking substrates, we have illustrated that draining air around $r_0$ can either completely eliminate or significantly reduce splashing. Correspondingly, on a smooth substrate without any pores and leakage, the air trapped at $r_0$ will initiate strong instability and lead to the onset of splashing. Therefore, we can quantitatively predict the splashing onset location on a smooth substrate, by finding $r_0$ with the intersection of $k_m^{-1}$ and $d$ curves, as shown in the inset of Fig.3d. Separately and independently, we can experimentally measure the location where splashing first appears, $r_{onset}$, with high speed photography. The measurements from the experiment, $r_{onset}$, and the predictions from our model, $r_0$, are directly compared in the main panel of Fig.3d: under extensive conditions with different liquids, velocities and substrates, the agreement between the experiment and the model is rather outstanding, providing a solid support for our picture of splashing generation.

## Discussion

With carefully designed porous substrates, we identify the KH-instability initiated within an ultra-thin air film as the origin of splashing on smooth surfaces. This picture agrees quantitatively with experimental verifications and illustrates the fundamental mechanism of splashing. However, we clarify that the KH-instability only provides a mechanism for the rim formation at the edge, which subsequently takes off; while the rupturing of the rim and liquid sheet may involve some other mechanism such as the Plateau-Rayleigh instability, which requires further investigation. We also note that our experiments are within the low-viscosity regime, where the surface tension dominates the viscous effect. This provides the ground for the application of KH-instability with surface tension only but without viscosity. For the splash of more viscous liquids, the viscous effect should be included and further study is required.

More interestingly, because such air entrapment occurs quite generally for liquid motion on solid substrates, our newly proposed instability may provide a novel mechanism for the common phenomena of liquid-solid wetting during dynamic motions: the growth of the instability within the ultra-thin air film may cause the initial touch between liquid and solid, which subsequently develops into the complete wetting. Further study along this direction may illustrate the ubiquitous dynamic-wetting process, and make a significant impact on the coating industry. This mechanism could also be crucial for impacts on super-hydrophobic surfaces [3, 4, 11, 12], where the air entrapment constantly occurs. The extension of KH-instability to the condition of an ultra-thin air film uncovers a new area for this classical instability analysis.

ACKNOWLEDGMENTS. We thank the great discussion and help from Michael Brenner, Hau Yung Lo, Qi Ouyang, Chu Zhang and Bo Zheng. This project is supported by the Hong Kong GRF grant CUHK404211, CUHK404912, and CUHK direct grant 4053081.